\documentclass[]{JHEP}
\usepackage{epsfig}
\usepackage{cite}

\title{CHARYBDIS: A Black Hole Event Generator}

\author{C.M.~Harris$^{\dagger}$, P.~Richardson$^{\ddagger}$ 
and B.R.~Webber$^{\dagger,\ddagger}$\\
$^{\dagger}$Cavendish Laboratory, University of Cambridge, Madingley Road, 
Cambridge, CB3~0HE, UK.\\
$^{\ddagger}$Theory Division, CERN, 1211 Geneva 23, Switzerland.\\
}

\abstract{{\small CHARYBDIS} is an event generator which simulates the production and decay of miniature black holes at hadronic colliders as might be possible in certain extra dimension models.  It interfaces via the Les Houches accord to general purpose Monte Carlo programs like {\small HERWIG} and {\small PYTHIA} which then perform the parton evolution and hadronization.  The event generator includes the extra-dimensional `grey-body' effects as well as the change in the temperature of the black hole as the decay progresses.  Various options for modelling the Planck-scale terminal decay are provided.}


\preprint{Cavendish-HEP-03/12 \\
		CERN-TH/2003-170}

\begin{document}

\section{Introduction}

Models with extra dimensions have become an area of much interest since the 
work of Arkani-Hamed, Dimopoulos and Dvali (ADD) \cite{ADD} and Randall and 
Sundrum (RS) \cite{RS}.  Like most models of physics beyond the Standard Model
they are seen as a more natural way of explaining the hierarchy problem, that 
is, why there are about sixteen orders of magnitude between the electroweak 
energy scale and the Planck scale at which gravity becomes large.  Such extra 
dimension models can also be motivated from string theory.

In extra dimension models, the usual 4-dimensional Planck scale is no longer 
considered to be a fundamental scale - instead it is derived from the 
fundamental D-dimensional Planck scale which can be as low as current 
experimental limits allow ($\sim$ 1 TeV).

If the fundamental Planck scale is of order a TeV, gravity is strong at such scales and cannot be ignored as is usually the case in particle physics.  The possibility then arises of particle accelerators at TeV-scale energies being able to produce miniature black holes.  These would then decay rapidly\footnote{This is only true in the ADD model, in the RS model the black holes can be stable on collider time scales\cite{Casadio:2001wh}.}
 by Hawking evaporation, giving rise to characteristic high-multiplicity final states.

There has already been much discussion in the literature on this issue, but 
little work has been done trying to realistically simulate black holes at
the Large Hadron Collider (LHC).  In this work we implement a simple model of
black hole production and decay which can be interfaced to existing Monte Carlo
programs using the Les Houches accord \cite{LHA}.  The major new theoretical input to the generator is the inclusion of the recently calculated `grey-body' factors for black holes in extra dimensions \cite{kmr1,kmr2,grey}.  We also take account of the recoil and change of temperature of the black hole during decay, and provide various models for the termination of the decay process.

\section{Black Hole Production and Decay}
\label{proddec}

The details of production and decay of black holes in extra dimension models 
are complicated and not particularly well understood.  Here we outline the 
theory and mention some of the assumptions which are usually made.

In theories with extra dimensions the $\sim$ TeV energy scale is considered
as fundamental - the 4D Planck scale ($M_{p(4)}\sim 10^{18}$ GeV) is then derived from it.  The relationship between the two energy scales is determined by the volume of the extra dimensions.  If $R$ is the size of all $n$ extra 
dimensions it can be shown, using Gauss' Law, that for $r\ll R$ then

\begin{equation}
V(r)\sim\frac{M}{M_p^{n+2}}\frac{1}{r^{n+1}},
\end{equation}

\noindent whereas for $r\gg R$

\begin{equation}
V(r)\sim\frac{M}{M_p^{n+2}R^n}\frac{1}{r}.
\end{equation}

In these expressions $M_p$ is the $(4+n)$-dimensional Planck mass (throughout this paper the conventions of \cite{DL} are used for $M_p$).  They show that the two energy scales are related (up to volume factors of order unity) by

\begin{equation}
M_{p(4)}^2\sim M_p^{n+2}R^n,
\end{equation}

\noindent which allows the sizes of extra dimensions to be calculated for 
different values of $n$ \cite{ADD}.  Short scale gravity experiments and particle collider experiments provide limits on the fundamental Planck scale.  However for the smaller values of $n$, the more stringent constraints come from astrophysical and cosmological data, albeit with larger uncertainties.  It is widely agreed that both $n=1$ and $n=2$ are ruled out by such data.  For a comprehensive recent review of these constraints see, for example, \cite{hewett}.

As the fundamental Planck scale is as low as $\sim$ TeV, it is possible for 
tiny black holes to be produced at the LHC when two partons pass within the 
horizon radius set by their centre-of-mass energy.  The black holes being 
considered in this work are in the $r\ll R$ regime, so an analogous approach 
to the usual 4D Schwarzschild calculation \cite{myers} shows the horizon 
radius for a non-spinning black hole to be

\begin{equation}
r_h = \frac{1}{\sqrt{\pi}M_p}\left(\frac{M_{BH}}{M_p}\right)^{\frac{1}{n+1}}\left(\frac{8\Gamma\left(\frac{n+3}{2}\right)}{n+2}\right)^{\frac{1}{n+1}},
\end{equation}

\noindent where $M_{BH}$ is the mass of the black hole.  

There has been much discussion in the literature (\emph{e.g.}~\cite{GT,vol1,
giddings,vol2,eardley}) about what the cross section for black hole production 
is, but the consensus opinion seems to be that the classical 
$\sigma\sim\pi r_h^2$ is valid (at least for black hole masses 
$M_{BH} \gg M_p$).  It is unclear for exactly what mass this cross section 
estimate starts to become unreliable, but for $M_{BH}$ close to the 
fundamental Planck scale a theory of quantum gravity would be required to 
determine the cross section.  The black holes produced may have any gauge and spin quantum numbers so to determine the the $p-p$ or $\overline{p}-p$ production cross section it is necessary to sum over all possible quark and gluon pairings.  Although the parton-level cross sections grows with black hole mass, the parton distribution functions (pdfs) fall rapidly at high energies and so the cross section also falls off quickly.

Once produced, these miniature black holes are expected to decay 
instantaneously on LHC detector time scales (typical lifetimes are 
$\sim 10^{-26}$ s).  The decay is made up of three major phases:

\begin{itemize}

\item
The balding phase in which the `hair' (asymmetry and moments due to the violent
production process) is lost;

\item
A Hawking evaporation \cite{hawking} phase (a brief spin-down phase during 
which angular momentum is shed from a Kerr black hole, then a longer 
Schwarzschild phase);

\item
A Planck phase at the end of the decay when the mass and / or the Hawking 
temperature approach the Planck scale.

\end{itemize}

It has been shown in \cite{emperan} that the majority of energy in Hawking 
radiation is emitted into modes on the brane (\emph{i.e.} as Standard Model 
particles) but a small amount is also emitted into modes in the bulk 
(\emph{i.e.} as gravitons).

In 4D the phase which accounts for the greatest proportion of the mass loss is the Schwarzschild phase \cite{page}.  A black hole of a particular mass is 
characterized by a Hawking temperature and as the decay progresses the black 
hole mass falls and the temperature rises.  It is assumed that a 
quasi-stationary approach to the decay is valid, that is the black hole has 
time to come into equilibrium at each new temperature before the next particle
is emitted.

For an uncharged, non-rotating black hole the decay spectrum is described by the following expression:

\begin{equation}
\label{spec}
 \frac{dN_{s,l,m}}{d\omega dt} = \frac{1}{2\pi}\frac{\Gamma_{s,l,m}}{exp[\omega/T_H]\mp1},
\end{equation}

\noindent where $s$ is the spin of the polarization degree of freedom being considered, $l$ and $m$ are angular momentum quantum numbers, and $\Gamma$ are the so-called `grey-body' factors.  The last term in the denominator is a spin statistics factor which is $-1$ for bosons and $+1$ for fermions.  The Hawking temperature in (\ref{spec}) is given by

\begin{equation}
T_H=\frac{n+1}{4\pi r_h}.
\end{equation}

\noindent Equation (\ref{spec}) can be used to determine the decay spectrum for
a particular particle \emph{e.g.} electrons.  Since there are two polarizations for spin-$\frac{1}{2}$ we obtain:

\begin{equation}
\frac{dN_{e^-}}{d\omega dt}=2 \sum_{l,m} \frac{dN_{1/2,l,m}}{d\omega dt}.
\end{equation}

\noindent  The expression for a particular flavour of quark would be identical but with an additional colour factor.  Slightly more care is required with massive gauge bosons since one of their degrees of freedom comes from the Higgs mechanism.  This means that, for example,

\begin{equation}
\frac{dN_{W^-}}{d\omega dt}=2 \sum_{l,m} \frac{dN_{1,l,m}}{d\omega dt}
+ \sum_{l,m} \frac{dN_{0,l,m}}{d\omega dt}.
\end{equation}

The grey-body factors modify the spectrum of emitted particles from that of a
perfect thermal black body \emph{even} in 4D \cite{hawking}.  They quantify the probability of transmission of the particles through the curved space-time outside the horizon, and can be determined from the absorption cross section for the emitted particle species.
For the Schwarzschild phase with $\omega\gg T_H$ geometric arguments show that $\Sigma_{l,m}\Gamma\propto(\omega r_h)^2$ in any number of dimensions, which means that at high energies the shape of the spectrum is like that of a black body.  However the low energy behaviour of the grey-body factors is spin-dependent and also depends on the number of dimensions.

In 4D it has long been known that for $s=0,\frac{1}{2} \mbox{ and }1$ the grey-body factors reduce
the low energy emission rate significantly below the geometrical optics value 
\cite{page, macG}.  The result is that both the flux and power spectra peak at 
higher energies than those for a black body at the same temperature.  The spin
dependence of the grey-body factors mean that they are necessary to determine 
the relative emissivities of different particle types from a black hole.  Until recently (see \cite{kmr1,kmr2,grey}) these have only been available in the literature for the 4D case \cite{page,sanchez}).  

The dependence both on energy and the number of dimensions means that the grey-body factors must be taken into account in any attempt to determine the number of extra dimensions by studying the energy spectrum of particles emitted from a black hole.  When studying black hole decay, other experimental variables may also be sensitive to these grey-body effects. 

Finally although the Planck phase cannot be properly understood without a full theory of quantum gravity, it is suggested that in this phase the black hole will decay to a few quanta with Planck-scale energies \cite{giddings}.

\section{Event Generator} 
\label{bhgen}

\subsection{Features of the event generator}

There are a number of features of the {\small CHARYBDIS} generator which, within the uncertainties of much of the theory, allow reliable simulation of black hole events.  Most notable is that unlike other generators (\emph{e.g.}~\cite{TRUENOIR}) the grey-body effects are fully included.  The generator also allows the black hole temperature to vary as the decay progresses and is designed for simulations with either $p-p$ or $\overline{p}-p$.

Due to the difficulty in modelling the balding phase and the lack of a full 
theory for quantum gravity to explain the Planck phase of the decay, the 
generator only attempts to model the Hawking evaporation phase (expected to 
account for the majority of the mass loss).  To provide a further simplification only non-spinning black holes are modelled.  This is perhaps a less good 
approximation but comparison with the 4D situation suggests that most of the 
angular momentum will be lost in a relatively short spin-down phase 
\cite{page2}.

It is possible that black hole decay does not conserve baryon number, for example by producing three quarks in a colour singlet.  However the treatment of processes which do not conserve baryon number in both the QCD evolution and hadronization is complicated and has only been studied for a few specific processes \cite{gibbs1,gibbs2,rpv,skands}.  At the same time the violation of baryon number is extremely difficult to detect experimentally and therefore the effect of including baryon number violation is not expected to be experimentally observable.  Therefore {\small CHARYBDIS} conserves baryon number in black hole production and decay.

\subsection{General description}

The black hole event generator developed attempts to model the theory as 
outlined in the previous section.  There are several related parameters and switches which can be set in the first part of the Les Houches subroutine \texttt{UPINIT} \cite{LHA}.  No other part of the charybdis1000.F code should be modified.

Firstly the properties of the beam particles must be specified.  \texttt{IDBMUP(1)} and \texttt{IDBMUP(2)} are their PDG codes (only protons and anti-protons are allowed) and the corresponding energies are \texttt{EBMUP(1)} and \texttt{EBMUP(2)}.  Note that these settings will over-write any in the main {\small HERWIG} or {\small PYTHIA} program files.

The geometric parton-level cross section, $\sigma=\pi r_h^2$, is used but the parameters \texttt{MINMSS} and \texttt{MAXMSS} allow the mass range for 
the black holes produced to be specified.  This means that the lower mass limit at 
which this expression for the cross section is thought to become valid can be 
adjusted.

Three other parameters which must be set before using the event generator are
\texttt{TOTDIM}, \texttt{MPLNCK} and \texttt{MSSDEF}.  The total number of dimensions in the model being used is given by \texttt{TOTDIM} (this must be set between 6 and 11).  There are a number of different definitions of the Planck mass (set using \texttt{MPLNCK}), but the parameter \texttt{MSSDEF} can be set to three different values to allow easy interchange between the three conventions outlined in Appendix A of \cite{GT}.  The conversions between these conventions are summarized in Table~\ref{massdefs}.

\TABULAR{|c|l|}{\hline
\texttt{MSSDEF} & Conversion\\
\hline
1 & $\mbox{\texttt{MPLNCK}}=(2^{n-2}\pi^{n-1})^{\frac{1}{n+2}}M_p$\\
2 & $\mbox{\texttt{MPLNCK}}=M_p$\\
3 & $\mbox{\texttt{MPLNCK}}=(2^{n-3}\pi^{n-1})^{\frac{1}{n+2}}M_p $\\
\hline}
{\label{massdefs}Definitions of the Planck mass}

It has been suggested that since black hole formation is a non-perturbative 
process, the momentum scale for evaluating the pdfs should be the inverse 
Schwarzschild radius rather than the black hole mass.  The switch 
\texttt{GTSCA} should be set to \texttt{.TRUE.} for the first of these options
and \texttt{.FALSE.} for the second.  It should be noted that, as confirmed in \cite{rizzo}, the cross sections quoted in reference \cite{GT} were actually calculated with the latter pdf scale.  The pdfs to be used are set using the Les Houches parameters \texttt{PDFGUP} and \texttt{PDFSUF}.

As discussed in section \ref{proddec}, the Hawking temperature of the black 
hole will increase as the decay progresses so that later emissions will 
typically be of higher energy.  However to allow comparison with other work 
which has ignored this effect, there is a switch \texttt{TIMVAR} which can be 
used to set the time variation of the Hawking temperature as on 
(\texttt{.TRUE.}) or off (\texttt{.FALSE.}).

The probabilities of emission of different types of particles are set according to the new theoretical results \cite{kmr1,kmr2,grey}.  Heavy particle production is allowed and can be controlled by setting the value of the \texttt{MSSDEC} parameter to \texttt{2} for top quark, W and Z, or \texttt{3} to include Higgs also (\texttt{MSSDEC=1} gives only light particles).  Heavy particle production spectra may be unreliable for choices of parameters for which the initial Hawking temperature is below the rest mass of the particle being considered.

If \texttt{GRYBDY} is set as \texttt{.TRUE.} the particle types and energies are chosen according to the grey-body modified emission probabilities and spectra.  If instead the \texttt{.FALSE.} option is selected, the black-body emission probabilities and spectra are used.  The choice of energy is made in the rest frame of the black hole before emission.  As overall charge must be conserved, when a charged particle is to be emitted the particle or anti-particle is chosen such that the magnitude of the black hole charge decreases.  This reproduces some of the features of the charge-dependent emission spectra in \cite{page} whilst at the same time making it easier for the event generator to ensure that charge is conserved for the full decay.

Although the Planck phase at the end of decay cannot be well modelled as it is
not well understood, the Monte Carlo event generator must have some way of
terminating the decay.  There are two different possibilities for this, each with a range of options for the terminal multiplicity.  

If \texttt{KINCUT=.TRUE.} termination occurs when the chosen energy for the emitted particle is ruled out by the kinematics of a two-body decay.  At this point an isotropic \texttt{NBODY} decay is performed on the black hole remnant where \texttt{NBODY} can be set between 2 and 5.  The \texttt{NBODY} particles are chosen according to the same probabilities used for the rest of the decay. The selection is then accepted if charge and baryon number are conserved, otherwise a new set of particles is picked for the \texttt{NBODY} decay. If this does not succeed in conserving charge and baryon number after \texttt{NHTRY} attempts the whole decay is rejected and a new one generated.  If the whole decay process fails for \texttt{MHTRY} attempts then the initial black hole state is rejected and a new one generated. 

In the alternative termination of the decay (\texttt{KINCUT=.FALSE.}), particles are emitted according to the energy  spectrum until $M_{BH}$ falls below \texttt{MPLNCK} and then an \texttt{NBODY} decay as described above is performed.  Any chosen energies which are kinematically forbidden are simply discarded. 

In order to perform the parton evolution and hadronization the general purpose event generators require a colour flow to be defined.  This colour flow is defined in the large number of colours ($N_c$) limit in which a quark can be considered as a colour line, an anti-quark as an anti-colour line and a gluon both a colour and anti-colour line.  A simple algorithm is used to connect all the lines into a consistent colour flow.  This algorithm starts with a colour line (from either a quark or a gluon) and then randomly connects this line with one of the unconnected anti-colour lines (either a gluon or an anti-quark).  If the selected partner is a gluon the procedure is repeated to find the partner for its colour line; if it is an anti-quark one of the other unconnected quark colour lines is selected.  If the starting particle was a gluon the colour line of the last parton is connected to the anti-colour line of the gluon.  Whilst there is no deep physical motivation for this algorithm it at least ensures that all the particles are colour-connected and the showering generator can proceed to evolve and hadronize the event.

After the black hole decay, parton-level information is written into the Les Houches common block \texttt{HEPEUP} to enable a general purpose event generator to fragment all emitted coloured particles into hadron jets, and generate all unstable particle decays (see below).

\subsection{Control switches, constants and options}

Those parameters discussed in the previous section which are designed to be set by the user are summarized in Table~\ref{parameters}.

\TABULAR{|c|l|c|c|}{\hline
Name & Description & Values & Default\\
\hline
\texttt{IDBMUP(2)} & {\small PDG} codes of beam particles & $\pm$2212 & 2212\\
\texttt{EBMUP(2)} & Energies of beam particles (GeV) & & 7000.0\\
\texttt{PDFGUP(2)} & {\small PDFLIB} codes for pdf author group & & $-1$\\
\texttt{PDFSUP(2)} & {\small PDFLIB} codes for pdf set & & $-1$\\  
\hline
\texttt{MINMSS} & Minimum mass of black holes (GeV) & $<\mbox{\texttt{MAXMSS}}$ & 5000.0\\
\texttt{MAXMSS} & Maximum mass of black holes (GeV) & $\leq\mbox{c.m. energy}$ & c.m. energy\\
\texttt{MPLNCK} & Planck mass (GeV) & $\leq\mbox{\texttt{MINMSS}}$ & 1000.0\\
\texttt{MSSDEF} & Convention for \texttt{MPLNCK} (see Table \ref{massdefs})& 1-3 & 2\\
\texttt{TOTDIM} & Total number of dimensions ($4+n$) & 6-11 & 6\\
\texttt{GTSCA} & Use $r_h^{-1}$as the pdf momentum scale & \texttt{LOGICAL} & \texttt{.FALSE.}\\
& rather than the black hole mass & &\\
\texttt{TIMVAR} & Allow $T_H$ to change with time & \texttt{LOGICAL}& \texttt{.TRUE.}\\
\texttt{MSSDEC} & Choice of decay products & 1-3 & 3\\
\texttt{GRYBDY} & Include grey-body effects & \texttt{LOGICAL} & \texttt{.TRUE.}\\
\texttt{KINCUT} & Use a kinematic cut-off on the decay & \texttt{LOGICAL} & \texttt{.FALSE.}\\
\texttt{NBODY} & Number of particles in remnant decay & 2-5 & 2\\
\hline}
{\label{parameters}List of parameters with brief descriptions, allowed values and default settings}

\subsection{Using charybdis1000.F}

The generator itself only performs the production and parton-level decay of the black hole. It is interfaced, via the Les Houches accord, to either {\small HERWIG} \cite{HERWIG,HERWIG2} or {\small PYTHIA} \cite{pythia} to perform the parton shower evolution, hadronization and particle decays.  This means that it is also necessary to have a Les Houches accord compliant version of either {\small HERWIG} or {\small PYTHIA} with both the dummy Les Houches routines (\texttt{UPINIT} and \texttt{UPEVNT}) and the dummy \texttt{PDFLIB} subroutines (\texttt{PDFSET} and \texttt{STRUCTM}) deleted.  For {\small HERWIG} the first Les Houches compliant version is {\small HERWIG6.500} \cite{HERWIG3}; for {\small PYTHIA} version 6.220 \cite{pythia2} or above is required.\footnote{Versions of {\small PYTHIA} above 6.200 support the Les Houches accord but can not handle more than 7 outgoing particles, which is necessary in black hole decays.}

The black hole code itself is available as a gzipped tar file at the web address \texttt{http://www.ippp.dur.ac.uk/montecarlo/leshouches/generators/charybdis/ }. The file includes the following code:

\begin{itemize}
\item{charybdis1000.F (code for the black hole generator)} 
\item{dummy.F (dummy routines needed if not using {\small PDFLIB})} 
\item{mainpythia.f (example main program for {\small PYTHIA})} 
\item{mainherwig.f (example main program for {\small HERWIG})} 
\item{charybdis1000.inc (include file for the black hole generator)} 
\end{itemize}

The general purpose event generator to be used must be specified in the Makefile (\emph{i.e.} \texttt{GENERATOR=HERWIG} or \texttt{GENERATOR=PYTHIA}) and also if {\small PDFLIB} is to be used (\texttt{PDFLIB=PDFLIB} if required, otherwise \texttt{PDFLIB=~}).  The name of the {\small HERWIG} or {\small PYTHIA} source and the location of the {\small PDFLIB} library must also be included. 

If the code is extracted to be run separately then the following should be taken into account:

\begin{itemize}
\item{charybdis1000.F will produce the {\small HERWIG} version by default when compiled, the flag \texttt{-DPYTHIA} should be added if the {\small PYTHIA} version is required;} 
\item{dummy.F will by default produce the version for use without \texttt{PDFLIB}, the flag \texttt{-DPDFLIB} should be added if \texttt{PDFLIB} is being used.} 
\end{itemize}

\subsection{List of subroutines}

Table~\ref{subs} contains a list of all the subroutines of the generator along with their functions.  Those labelled by {\small HW/PY} are 
{\small HERWIG / PYTHIA} dependent and are pre-processed according to the \texttt{GENERATOR} flag in the Makefile.  Many of the utility routines are identical to routines which appear in the {\small HERWIG} program.

\TABULAR{|l|l|}{\hline
Name & Description\\
\hline
& Les Houches routines\\
\hline
\texttt{UPINIT} & Initialization routine\\
\texttt{UPEVNT} & Event routine\\
\hline
& Particle decays\\
\hline
\texttt{CHDFIV} & Generates a five-body decay\\
\texttt{CHDFOR} & Generates a four-body decay\\
\texttt{CHDTHR} & Generates a three-body decay\\
\texttt{CHDTWO} & Generates a two-body decay\\
\hline
& Hard subprocess and related routines\\
\hline
\texttt{CHEVNT} & Main routine for black hole hard subprocess\\
\texttt{CHFCHG} & Returns charge of a SM particle\\
\texttt{CHFMAS} & Returns mass of a SM particle ({\small HW/PY})\\
\texttt{CHHBH1} & Chooses next particle type if \texttt{MSSDEC=1}\\
\texttt{CHHBH2} & Chooses next particle type if \texttt{MSSDEC=2}\\
\texttt{CHHBH3} & Chooses next particle type if \texttt{MSSDEC=3}\\
\texttt{CHPDF} & Calculates the pdfs ({\small HW/PY})\\
\hline
& Random number generators\\
\hline
\texttt{CHRAZM} & Randomly rotates a 2-vector\\
\texttt{CHRGEN} & Random number generator ({\small HW/PY})\\
\texttt{CHRLOG} & Random logical\\
\texttt{CHRUNI} & Random number: uniform\\
\hline
& Miscellaneous utilities\\
\hline
\texttt{CHUBHS} & Chooses particle energy from spectrum\\
\texttt{CHULB4} & Boost: rest frame to lab, no masses assumed\\
\texttt{CHULOB} & Lorentz transformation: rest frame$\rightarrow$lab\\
\texttt{CHUMAS} & Puts mass in 5th component of vector\\
\texttt{CHUPCM} & Centre-of-mass momentum\\
\texttt{CHUROB} & Rotation by inverse of matrix $R$\\
\texttt{CHUROT} & Rotation by matrix $R$\\
\texttt{CHUSQR} & Square root with sign retention\\
\texttt{CHUTAB} & Interpolates in a table\\
\hline
& Vector manipulation\\
\hline
\texttt{CHVDIF} & Vector difference\\
\texttt{CHVEQU} & Vector equality\\
\texttt{CHVSUM} & Vector sum\\
\hline}
{\label{subs}List of subroutines with brief descriptions}

\subsection{Sample plots}

Figures \ref{higgs}-\ref{photon} show the results, at parton level, of neglecting the time variation of the black hole temperature (\texttt{TIMVAR=.FALSE.}, dashed line) or the grey-body factors (\texttt{GRYBDY=.FALSE.}, dot-dashed line) for initial black hole masses in the range from \texttt{MINMSS=5000.0} to \texttt{MAXMSS=5500.0}, with the default values for the other parameters.  The solid line is for simulations with the default parameter settings (but with the same reduced range of initial black hole masses used in the other two cases).

The effect of time variation is to harden the spectra of all particle species.
However, the effect of the grey-body factors depends on the spin, in this case slightly softening the spectra of scalars and fermions but hardening the spectrum of gauge bosons.

\FIGURE{\epsfig{file=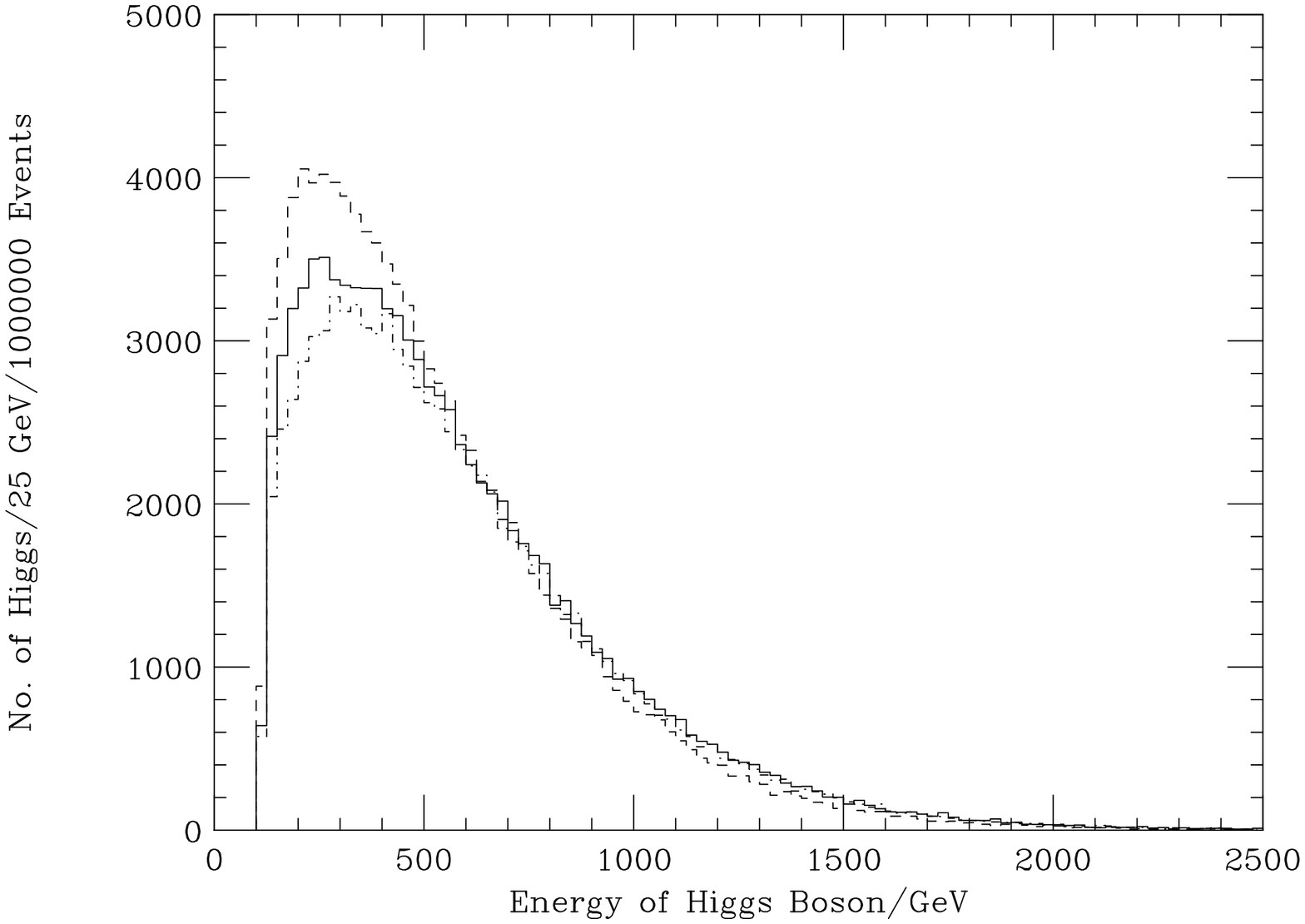, angle=90, width=.85\textwidth}
\caption{Parton-level energy spectra of Higgs bosons, $m_H = 115$~GeV. Solid: predicted energy spectrum of Higgs bosons from decay of black holes with initial masses 5.0-5.5 TeV.  Dashed: neglecting time variation of temperature.  Dot-dashed: neglecting grey-body factors.}
\label{higgs}}

\FIGURE{\epsfig{file=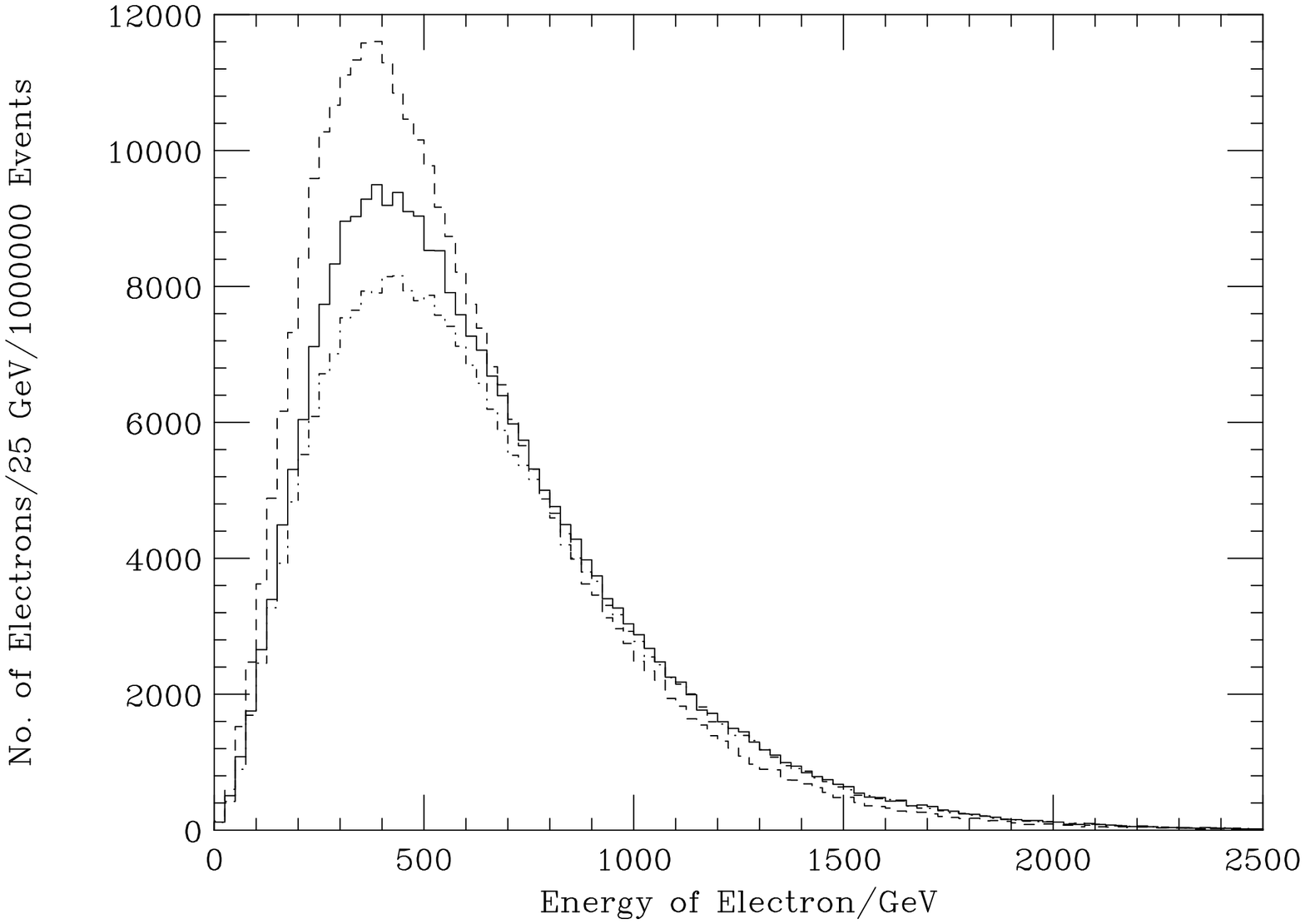, angle=90, width=.85\textwidth}
\caption{Parton-level energy spectra of electrons and positrons.  As Figure \ref{higgs} but for electron and positron spectra.}
\label{electron}}

\FIGURE{\epsfig{file=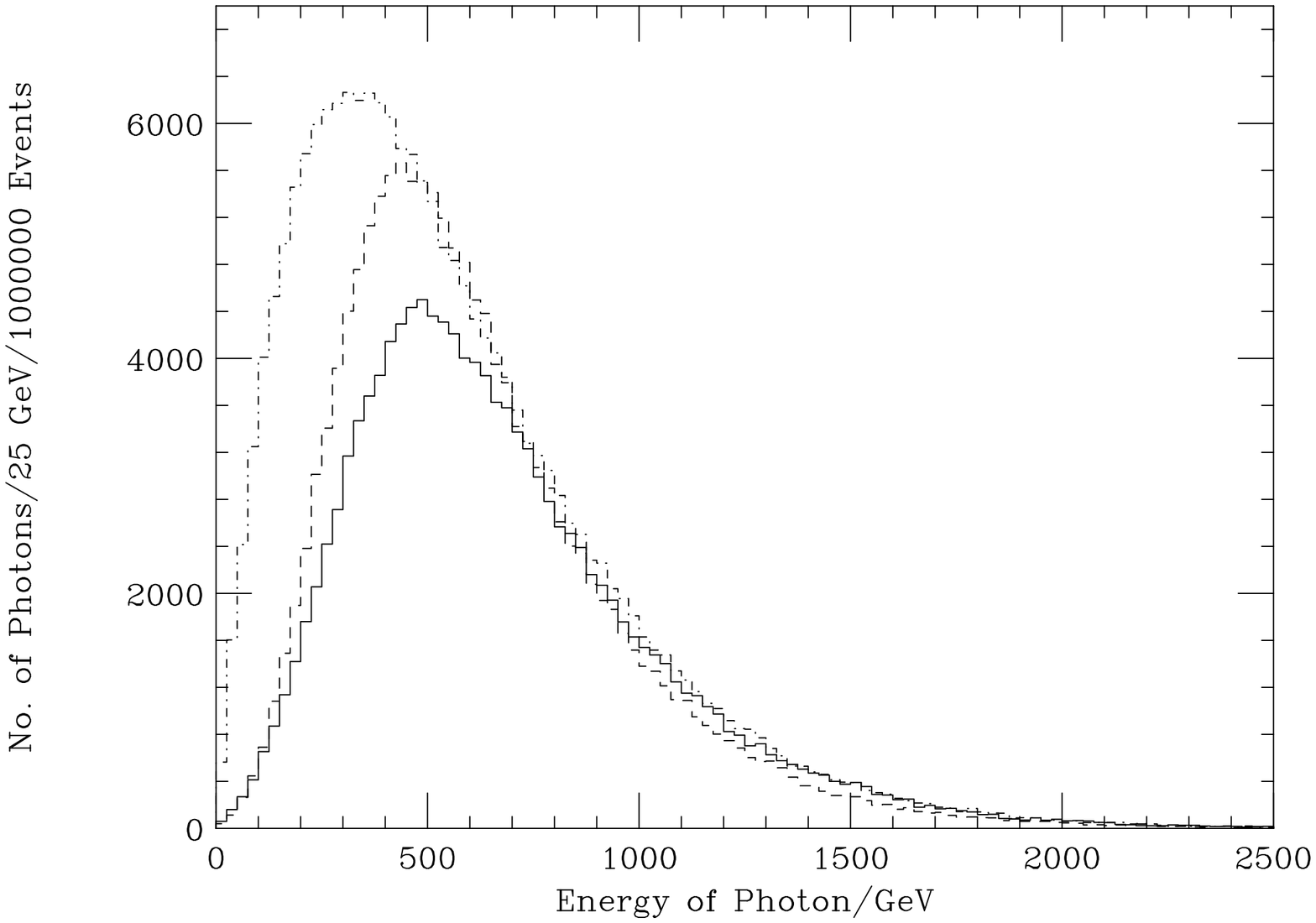, angle=90, width=.8\textwidth}
\caption{Parton-level energy spectra of photons.  As Figure \ref{higgs} but for photon spectra.}
\label{photon}}

Results of a fuller study of signatures of black hole production and decay at the LHC will be presented elsewhere \cite{ali}.

\acknowledgments

We thank members of the Cambridge SUSY Working Group, members of the ATLAS collaboration, and also P.~Kanti and J.~March-Russell for helpful discussions.  Thanks also to the authors of {\small HERWIG} for the code incorporated into this generator, and to  T.~Sj\"ostrand for help with running {\small CHARYBDIS} with {\small PYTHIA}.  This work was funded by the U.K. Particle Physics and Astronomy Research Council.

\end{document}